\documentclass[aip,jcp,preprint,amsmath,amssymb]{revtex4-2}
\usepackage{graphicx}
\usepackage[suffix=]{epstopdf}
\usepackage{natmove}

\begin{document}
\title{Effects of Ligand vs.~Linker on Phase Behavior and Mechanical Properties of Nanoparticle Gels}

\author{Qizan Chen}
\affiliation{Artie McFerrin Department of Chemical Engineering, Texas A{\upshape\&}M University, College Station, TX 77843, United States}

\author{Dinesh Sundaravadivelu Devarajan}
\affiliation{Artie McFerrin Department of Chemical Engineering, Texas A{\upshape\&}M University, College Station, TX 77843, United States}

\author{Arash Nikoubashman}
\affiliation{Leibniz-Institut f{\"u}r Polymerforschung Dresden e.V., Hohe Stra{\ss}e 6, 01069 Dresden, Germany}
\affiliation{Institut f{\"u}r Theoretische Physik, Technische Universit{\"a}t Dresden, 01069 Dresden, Germany}

\author{Michael P. Howard}
\email{mphoward@auburn.edu}
\affiliation{Department of Chemical Engineering, Auburn University, Auburn, Alabama 36849, United States}

\author{Jeetain Mittal}
\email{jeetain@tamu.edu}
\affiliation{Artie McFerrin Department of Chemical Engineering, Texas A{\upshape\&}M University, College Station, TX 77843, United States}
\affiliation{Department of Chemistry, Texas A{\upshape\&}M University, College Station, TX 77843, United States}
\affiliation{Interdisciplinary Graduate Program in Genetics and Genomics, Texas A{\upshape\&}M University, College Station, TX 77843, United States}

\begin{abstract}
Nanoparticle gels have attracted considerable attention due to their highly tunable properties. One strategy for producing nanoparticle gels involves using strong local attractions between polymeric molecules, such as DNA hybridization or dynamic covalent chemistry, to form percolated nanoparticle networks. These molecules can be used in two distinct roles: as ``ligands'' with  one end grafted to a nanoparticle or as ``linkers'' with both ends free. Here, we explore how these roles shape the phase behavior and mechanical properties of gel-like nanoparticle assemblies using coarse-grained simulations. We systematically vary the interaction strength and bending stiffness of both ligands and linkers. We find that phase separation can be limited to low nanoparticle volume fractions by making the ligands rigid, consistent with previous studies on linked nanoparticle gels. At fixed interaction strength and volume fraction, both ligand- and linker-mediated nanoparticle assemblies show similar mechanical responses as bending stiffness is varied. However, a comparison between the two association schemes reveals that the linked nanoparticles form rigid percolated networks that are less stretchable than the ligand-grafted gels, despite exhibiting similar tensile strength. We attribute these differences between ligands and linkers to the distinct structural arrangement of nanoparticles within the gel. Our findings highlight the potential to use different association schemes to tune specific mechanical properties.
\end{abstract}

\maketitle

\section{Introduction} \label{Sec1}
Nanoparticle gels are promising functional materials because their properties can be modulated through their mesoscopic structure, which is determined by the physical or chemical interactions between the nanoparticles.\cite{SaezCabezas_PNAS2018, CKnorowski_COSSMS2011, VladimirLesnyak_ACSNano2010, YoungSun_StatMechApps2005, RobertJMacfarlane_PNAS2009, AllisonMGreen_NanoLett2022, MurariSingh_JPCC2022, Bassani_ACSNano2024} For example, the solid-like network structure of gels can give rise to distinct mechanical properties \cite{JakeSong_ACSNano2020} as well as optical properties facilitated by coupling effects such as local surface plasmon resonance.\cite{NaomiJHalas_ChemRev2011} Compared to nanoparticle superlattices,\mbox{\cite{MichaelABoles_ChemRev2016, ZacharyMSherman_AccChemRes2021, MichaelBRoss_PNAS2015}} nanoparticle gels provide a potentially more robust, processable alternative. Previous experimental and simulation studies have also shown that modulating the mesostructure within nanoparticle gels provides ample opportunities to tune their properties.\cite{ManuelNDominguez_CoM2020, IndikaUArachige_AccChemRes2007, ChristophZiegler_MIA2017}

Nanoparticle gels are typically made by inducing a short-range attraction between the nanoparticles.\cite{FabianMatter_NanoToday2020} Various strategies can be used to create attraction, including solvent-mediated van der Waals attraction and polymer-induced depletion attraction.\cite{ManosAnyfantakis_SoftMatter2009, SaezCabezas_PNAS2018} The attraction between nanoparticles causes the fluid-like nanoparticle dispersion to become thermodynamically unstable, then the nanoparticles kinetically arrest as a gel during their spinodal decomposition into dilute and condensed phases.\cite{AaronPREberle_PRL2011} This mechanism produces a ``nonequilibrium'' gel that is susceptible to aging because the arrested phase separation can slowly continue over time.\cite{PeterJLu_Nature2008, EmanuelaZaccarelli_JPCM2007, JoepRouwhorst_NatComm2020}

Alternatively, nanoparticles with a limited number of highly localized attractive sites (so-called ``patchy particles'') can form long-lived networks without phase separation, resulting in an ``equilibrium'' gel.\cite{DanieldelasHeras_JCP2011, BALindquist_JCP2016, PICTeixeira_COCIS2017, EmanuelaZaccarelli_COCIS2017, MichaelPHoward_JCP2019, Ilnytskyi_softmat2018} Because such networks are thermodynamically stable, equilibrium gels tend to be less susceptible to aging than nonequilibrium gels. Equilibrium gels have been formed in simulations of patchy nanoparticles by directly limiting the number of attractive patches\cite{EmanuelaBianchi_PRL2006, JohnRusso_JCP2009, 07Bigall2009Hydrogels} and in experiments using trivalent and tetravalent DNA nanostars.\cite{SilviaBiffi_PNAS2013, LorenzoRovigatti_ACSNano2014, JFernandezCastanon_JCP2018} However, these types of valence limitation require careful nanoscopic control that presents challenges for large-scale fabrication and tunability. One way to overcome these challenges is by introducing a secondary ``linking'' species,  such as polymers,\cite{JieChen_JCP2015,JunhuaLuo_SoftMatter2015,ChuanzhuangZhao_SoftMatter2012} DNA,\cite{YugangZhang_NatNano2013, HuimingXiong_JACS2008} and ionic molecules,\cite{VladimirLesnyak_ACSNano2010, AmitaSingh_AngewChemie2015, VladimirSayevich_AngewChemie2016} whose concentration can be used to restrict valence macroscopically.\cite{BALindquist_JCP2016, ManuelNDominguez_CoM2020} For example, recent simulations indicated that adjusting the number ratio of linkers to nanoparticles and modifying the flexibility of oligomeric linkers both significantly affected the phase behavior and structure of the resulting gels.\cite{MichaelPHoward_JCP2019, ZacharyMSherman_AccChemRes2021, MichaelPHoward_JCP2021}

For the strategies described above, both ends of the linking molecules are not permanently attached to the nanoparticle. Alternatively, one end of a molecule can be grafted to a nanoparticle to act as a surface-bound ligand that facilitates attraction. For example, reversible gel assemblies have been created using ligands with dynamic covalent chemistry strategies,\cite{StuartJRowan_AngewChemie2002,YinghuaJin_ChemSocRev2013,ManuelNDominguez_CoM2020} and there is growing interest in reversible assemblies involving DNA-grafted nanoparticles.\cite{ChadAMirkin_Nature1996, CKnorowski_COSSMS2011, WBenjaminRogers_NatRevMat2016, ZhenlongLi_ACSApplBioMat2021, QiAn_PolyChem2019} While strategies for using linker design to control gelation have been relatively well-studied,\cite{MichaelPHoward_JCP2019, MichaelPHoward_JCP2021, MurariSingh_JPCC2022, BALindquist_JCP2016, Bassani_ACSNano2024} less is known about how ligand design modulates the phase behavior and mechanical properties of gels and to what extent ligand-mediated and linker-mediated gelation differ.

In this work, we employ coarse-grained molecular dynamics simulations of generic nanoparticles and linear polymers to investigate how the association mechanism, where the polymers act either as ligands or linkers, influences the phase behavior and mechanical properties of nanoparticle gels. Furthermore, we systematically vary the polymer stiffness, finding that the range of volume fractions over which phase separation occurs gradually narrows with increasing stiffness, bearing similarities to linked nanoparticle gels. Under uniaxial tensile deformation, equilibrium gels formed through both ligands and linkers showed noticeable changes in mechanical properties such as their shear modulus, tensile strength, and network stretchability as the ligand stiffness varied. Notably, a comparison between ligand-grafted nanoparticle gels and linked nanoparticle gels revealed distinct differences in certain mechanical properties, highlighting the ability to tune the mechanical response of equilibrium nanoparticle gels solely through the association scheme.

\section{Model and Simulation Methods} \label{Sec2}
\begin{figure}
\centering
\includegraphics{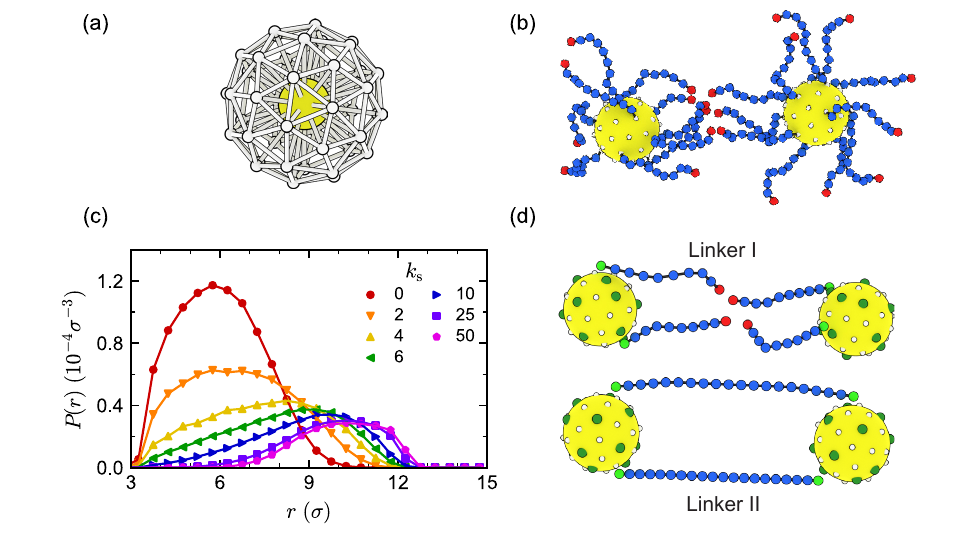}
\caption{(a) Schematic of bare nanoparticle (yellow, not drawn to scale) and 42 potential grafting sites (white), along with the bonds holding their shape. (b) Schematic of two interacting ligand-grafted nanoparticles. The terminal beads (red) facilitate ligand--ligand attraction. (c) Probability density function $P$ to observe the terminal beads of the ligands at a distance $r$ from the center of the bare nanoparticle as a function of polymer bending stiffness $k_{\mathrm{s}}$. The distribution is normalized such that $\int4\pi\,r^{2}P(r)\,\mathrm{d}r=1$. (d) Schematic of the two linker models that were compared to ligands. For Linker I, the linker-attaching terminal beads (red) facilitate linker--linker attraction (controlled by $\varepsilon_\mathrm{l}$), as for the ligands, but the nanoparticle-attaching terminal beads (green) facilitate linker--nanoparticle attraction (controlled by $\varepsilon_\mathrm{c}$). For Linker II, the nanoparticle-attaching terminal beads (green) facilitate only linker--nanoparticle attraction.}
\label{figure1}
\end{figure}

Similarly to previous work on DNA-functionalized nanoparticles,\cite{YajunDing_JCP2014} we used a coarse-grained model of a spherical nanoparticle with surface-grafted ligands. The bare nanoparticle had diameter $d_{\rm c}=6\,\sigma$, where $\sigma$ is the units of length in our simulations. We created grafting sites for the ligands by subdividing the faces of an icosahedron once and scaling the resulting 42 vertices to lie on the surface of the bare nanoparticle [Fig.~\ref{figure1}(a)]. These vertices were explicitly represented as particles with mass $m$ and diameter $\sigma$, where $m$ is the unit of mass in our simulations. We then iteratively grafted $14$ linear polymers onto the vertex sites, starting from an initial randomly chosen site that was the same for all nanoparticles, followed by a site that is farthest from the previous site(s) [Fig.~\ref{figure1}(b)]. Each polymer consisted of 10 beads with mass $m$ and diameter $d_{\rm l} = \sigma$. In total, the nanoparticle was represented by one central particle used to compute the bare nanoparticle interactions that also had mass $m$, 42 grafting sites, and 140 polymer beads, so the total mass of a ligand-grafted nanoparticle was $183\,m$. We used $N_{\rm c}=1000$ nanoparticles and varied the nanoparticle volume fraction $\eta=N_{\rm c}\pi d_{\rm c}^3/(6V)$ between 0.01 and 0.13 in steps of 0.02 by adjusting the volume $V$ of the simulation box. All simulations were performed at constant temperature $T$.

We modeled the excluded-volume interactions between the bare nanoparticles and between the bare nanoparticles and polymer beads using the purely repulsive Weeks--Chandler--Andersen (WCA) potential,\cite{JohnDWeeks_JCP1971}
\begin{equation}\label{eq:wca}
{\beta}U_{\mathrm{WCA}}(r)=\begin{cases}
    \displaystyle 4\left[\left(\frac{d_{ij}}{r}\right)^{12}-\left(\frac{d_{ij}}{r}\right)^6\right]+ 1, & r \leq 2^{1 / 6} d_{ij} \\
    0, & r>2^{1 / 6} d_{ij}
\end{cases}
,
\end{equation}
where $\beta = 1/(k_{\rm B} T)$, $k_{\rm B}$ is the Boltzmann constant, $r$ is the distance between particles, and $d_{ij} = (d_i + d_j)/2$ is the average diameter of the two interacting particles having types $i$ and $j$. We maintained the arrangement of grafting sites by connecting them to their nearest neighbors and to the bare nanoparticle using a stiff harmonic potential,
\begin{gather}\label{eq:harmonic}
\beta U_{\mathrm{h}}(r)=\frac{k_{\mathrm{h}}}{2}(r-r_{\mathrm{h}})^2,
\end{gather}
where the spring constant is $k_{\mathrm{h}} = 5000\,\sigma^{-2}$ and the rest length $r_{\mathrm{h}}$ is set for each bond using the distance in the initial configuration. Bonds between beads within the polymer chains were modeled using the finitely extensible nonlinear elastic potential,\cite{HaroldRWarner_IECF1972, MarvinBishop_JCP1979} 
\begin{gather}\label{eq:fene}
\beta U_{\mathrm{f}}(r)=-\frac{k_{\mathrm{f}}r_{\mathrm{f}}^2}{2}\ln\left [1-\left(\frac{r}{r_{\mathrm{f}}}\right)^2 \right],
\end{gather}
with spring constant $k_{\mathrm{f}}=30\,\sigma^{-2}$ and maximum bond extension $r_{\mathrm{f}}=1.5\,\sigma$. The terminal bead of a ligand was grafted to the nanoparticle using Eqs.~\eqref{eq:wca} and \eqref{eq:fene}.

The polymer flexibility was controlled using a harmonic bending potential, applied between three consecutively bonded polymer beads,
\begin{equation}\label{eq:stiffness}
\beta U_{\mathrm{s}}(\theta) = \frac{k_{\mathrm{s}}}{2}\left(\theta-\pi\right)^2,
\end{equation}
where $\theta$ is the angle between the beads and $k_{\mathrm{s}}$ is the spring constant that creates bending stiffness. To classify the polymer as flexible, semiflexible, and rigid, we varied $k_{\mathrm{s}}$ and computed the probability density function $P$ for the distance $r$ between the terminal bead of the ligand and the center of the bare nanoparticle for a single ligand-grafted nanoparticle [Fig.~\ref{figure1}(c)]. We found that the maximum in $P(r)$ shifted significantly toward larger $r$ as $k_{\mathrm{s}}$ increased, with the most pronounced changes occurring up to about $k_\text{s} \approx 6$. Since the persistence length in our model increases almost linearly with $k_{\mathrm{s}}$,\cite{Nikoubashman_JCP2016} this behavior is in line with the expectation that an initially flexible chain will become semiflexible when its persistence length is comparable to its contour length and rigid when its persistence length exceeds its contour length. Based on our measurements of $P(r)$, we defined polymers with $k_{\mathrm{s}}= 0, 6,\,\textrm{and}\,50$ as flexible, semiflexible, and rigid, respectively.

Attraction between the terminal beads of the ligands on different nanoparticles was modeled using the Lennard-Jones (LJ) potential,
\begin{equation}\label{eq:lj}
    \beta U_{\mathrm{LJ}}(r)=4 \varepsilon_{\mathrm{l}}\left[\left(\frac{d_{\rm l}}{r}\right)^{12}-\left(\frac{d_{\rm l}}{r}\right)^6\right],
\end{equation}
where $\varepsilon_{\rm l}$ sets the strength of the attraction. Following common practice, this potential was truncated and shifted to zero at $r = 3\,d_{\rm l}$ to reduce computational costs. We excluded attractions between ligands within the same nanoparticle to prevent ``loop'' formation\cite{MichaelPHoward_JCP2019, MichaelPHoward_JCP2021} that tends to inhibit network formation. Such a model is similar to a double-stranded DNA-grafted nanoparticle, for which the interactions between DNA chains bound to the same nanoparticle can be restricted using their rigidity and by designing noncomplementary binding sequences.\cite{SungYongPark_Nature2008, RunfangMao_PNAS2023}

To compare ligand-mediated gelation with linker-mediated gelation, we created two complementary linker models based on our ligand model [Fig.~\ref{figure1}(d)]. Linker I had the same degree of polymerization as the ligands (10 beads), but featured a terminal bead attracted to the grafting sites on the bare nanoparticle. Linker II contained twice as many monomers as Linker I (20 beads) and had only terminal beads capable of connecting to the grafting sites on the nanoparticles. Hence, Linker I can be regarded as a ligand without a permanent graft to the nanoparticle, while Linker II effectively behaves as two Linker I molecules that are permanently and exclusively paired. The number of linker molecules was chosen to keep the total number of monomers the same in all simulations. All three association schemes (ligand, Linker I \& II) can produce a comparable chain-like connection between two nanoparticles; however, different bonding motifs are possible for each model, and the mechanical properties associated with the connections may also differ. The attraction between the linker end(s) and a grafting site was modeled using the LJ potential [Eq.~\eqref{eq:lj}] with attraction strength $\varepsilon_{\rm c}$; the rest of the linker interactions were the same as for the ligands.
 
We performed molecular dynamics simulations using HOOMD-blue (version 2.9.3)\cite{JoshuaAAnderson_CMS2020} with features extended using azplugins (version 0.10.1).\cite{AZplugins} The integration time step was 0.005 $\tau$, and we maintained the temperature $T$ using a Langevin thermostat with friction coefficient 0.01 $m/\tau$, where $\tau=\sqrt{m\sigma^2 \beta}$ is the unit of time in the simulations. Following previous studies,\cite{MichaelPHoward_JCP2019, MurariSingh_JPCC2022} we prepared configurations under different conditions by slow annealing: First, we created three different initial configurations by dispersing the $N_{\rm c} = 1000$ nanoparticles randomly in a large cubic simulation box with edge length $L =1781\,\sigma$ and periodic boundary conditions. Initially, there were no attractions between ligands [\textit{i.e.}, only purely repulsive interactions through Eq.~\eqref{eq:wca}], and the box was compressed to the desired volume fraction $\eta$ by reducing the edge length at a constant rate over a $5\times{10}^3$\,$\tau$ time period, followed by an equilibration period of $5\times{10}^4\,\tau$ at the required volume fraction. We then gradually introduced attraction between the ligands by increasing $\varepsilon_\mathrm{l}$ to 5.0 in steps of 0.5, with each $\varepsilon_\mathrm{l}$ value being simulated for a duration of ${10}^5\,\tau$. We confirmed that for the initial purely repulsive interactions, both nanoparticles and polymers were able to diffuse at least the nanoparticle radius within $100\,\tau$ \cite{freud2020}, allowing the system to relax and explore new configurations (Figs. S1 and S2).

\begin{figure}
\centering
\includegraphics{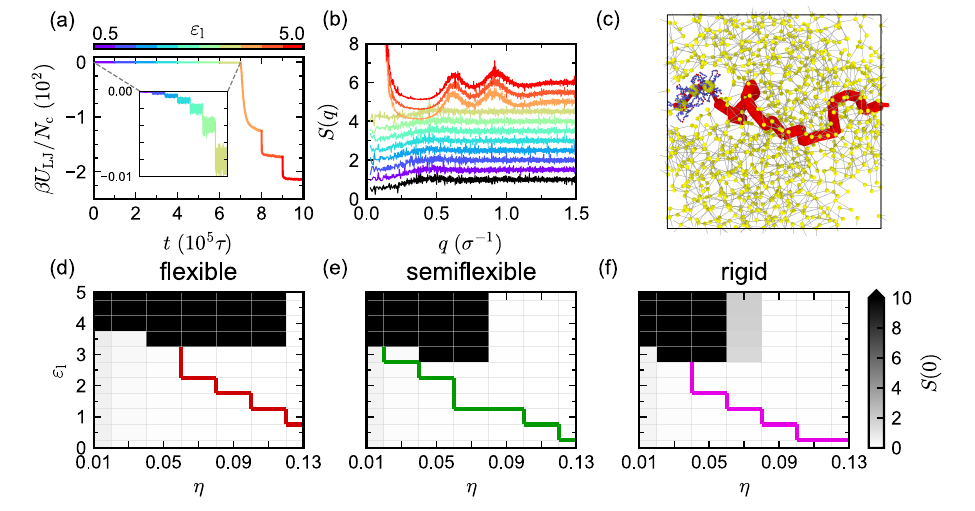}
\caption{(a) Time evolution of the attractive potential energy per nanoparticle $\beta U_{\mathrm{LJ}}/N_{\rm c}$ and (b) nanoparticle partial static structure factor $S(q)$ for flexible ligands with varied ligand--ligand attraction strengths $\varepsilon_\mathrm{l}$ when $\eta=0.01$. The inset in (a) shows a zoomed-in view of the potential energy before the large change associated with phase separation. In (b), the black line corresponds to no attraction between the ligands, and each $S\left(q\right)$ curve is vertically offset by 1 from the one below for clarity. (c) Illustration of a percolated path between nanoparticles for flexible ligands with $\varepsilon_\mathrm{l}=5$ when $\eta=0.13$. (d--f) $S(0)$ and percolation line for (d) flexible, (e) semiflexible, and (f) rigid ligands as a function of $\eta$ and $\varepsilon_\mathrm{l}$.}
\label{figure2}
\end{figure}

An example of the evolution of the attractive potential energy per nanoparticle $\beta U_{\mathrm{LJ}} / N_{\rm c}$ during the annealing process is shown in Fig.~\ref{figure2}(a) for $\eta = 0.01$ and $k_{\mathrm{s}}=0$. The final configurations obtained at each $\varepsilon_\mathrm{l}$ value were then used to perform an additional simulation for $10^4\,\tau$ at each value of $\varepsilon_{\rm l}$, from which 1000 configurations were collected at uniform time intervals for characterizing the structure and phase behavior. A similar procedure was used for preparing the linker-mediated assemblies, in which we incrementally increased $\varepsilon_\mathrm{c}$ simultaneously with $\varepsilon_\mathrm{l}$ from 0 to 5. We then additionally increased $\varepsilon_\mathrm{c}$ to 10, with a time period of ${10}^5\,\tau$ after the increase. We note that all reported averages are computed over multiple independent simulations and the configurations collected from each.

\section{Results and Discussion} \label{Sec3}
\subsection{Ligand-grafted nanoparticle assemblies} \label{Sec3.1}
Following previous studies,\cite{MichaelPHoward_JCP2019, MichaelPHoward_JCP2021} we first investigated the mesostructure of the ligand-grafted nanoparticle assemblies. To characterize whether the nanoparticles formed a single phase or tended toward phase separation, we computed their partial static structure factor,
\begin{gather}\label{e6}
S(\mathbf{q})=\frac{1}{N_\text{c}}\sum_{j,k}^{N_\mathrm{c}} e^{-i \mathbf{q} \cdot\left(\mathbf{r}_j-\mathbf{r}_k\right)},
\end{gather}
where $\mathbf{q}=2\pi\mathbf{n}/L$ is the wavevector, $\mathbf{n}$ is a vector of integers, and $\mathbf{r}_j$ is the position of the $j$-th nanoparticle. We averaged $S(\mathbf{q})$ over equivalent wavevector magnitudes $q = |\mathbf{q}|$ and the 1000 configurations collected from each of our three independent simulations to obtain an isotropically averaged $S(q)$ [Fig.~\ref{figure2}(b)]. A significant increase in $S$ as $q \to 0$ can be used to demarcate the region in which the nanoparticle dispersion tends toward separation into dilute and condensed phases. To this end, we extrapolated $S(0) \equiv S(q\to 0)$ by fitting a Lorentzian functional form to $S$ for the 22 smallest $q$ we calculated.\cite{MichaelPHoward_JCP2019, MichaelPHoward_JCP2021} (Note that the ``$-$'' in the Lorentzian functional form in these references is a typographical error and should be a ``$+$''.) Based on previous work,\cite{MichaelPHoward_JCP2019, MichaelPHoward_JCP2021} we considered $S(0)>10$ as indicative of phase separation.

We additionally evaluated the percolation of the nanoparticles into a space-spanning network. Specifically, we considered two nanoparticles to be connected if the distance between any attractive particles in their ligands was less than $1.4\sigma$, which corresponds to the approximate location of the first minimum in the radial distribution function between attractive particles of ligands on different nanoparticles (Fig.~S3). We considered a nanoparticle assembly to be percolated if a continuous path could be traced between any nanoparticle in the central simulation box and its periodic images in all three Cartesian directions [Fig.~\ref{figure2}(c)]. These calculations were performed using NetworkX.\cite{AricAHagberg_PPSC2008} We then defined a percolation boundary that demarcated the conditions under which more than $90\%$ of sampled configurations were percolated.

The extrapolated $S(0)$ and percolation lines are shown in Figs.~\ref{figure2}(d)--(f) for flexible, semiflexible, and rigid ligands. Similar diagrams for other bending stiffnesses are available in Fig.~S4. We found that $S(0) < 10$, indicating that the nanoparticles remained in a single phase, when $\varepsilon_{\rm l} \lesssim 3$ regardless of $\eta$, although percolated networks still formed for sufficiently large $\eta$. For $\varepsilon_{\rm l} \gtrsim 3$, we found $S(0) > 10$ over a limited range of $\eta$, indicating phase separation of the nanoparticles. This phase-separated region narrowed with increasing bending stiffness, extending until $\eta \approx 0.13,\,0.07,\,\textrm{and}\,0.05$ for the flexible, semiflexible, and rigid ligands, respectively. This behavior contrasts with previous observations for linker-mediated nanoparticle assemblies,\cite{MichaelPHoward_JCP2021} where both flexible and rigid linkers led to phase separation, whereas semiflexible linkers maintained a stable fluid phase due to their tendency to bind both ends to the same nanoparticle in ``loop'' motifs. Our model explicitly prohibits loop formation, supporting the hypothesis of prior works that loop formation may suppress phase separation.\cite{MichaelPHoward_JCP2019, MichaelPHoward_JCP2021}

We selected $\eta=0.13$ and $3 \le \varepsilon_\mathrm{l} \le 5$ for further investigation of the mechanical properties of the nanoparticle assemblies because these conditions produced percolated nanoparticle assemblies that were not phase separated for all bending stiffnesses investigated. For each of our three independent simulations, we subjected the final assemblies from the annealing to uniaxial tensile deformation along each of the three Cartesian directions, giving a total of 9 measurements for each ligand. We extended the length of the simulation box in the chosen direction by a factor $\lambda$ that increased at a constant rate of $\dot\lambda=5\times{10}^{-4}\,\tau^{-1}$. To conserve the system volume, the lengths of the simulation box in the other two directions were simultaneously compressed by a factor of $\lambda^{1/2}$, consistent with a Poisson's ratio of 0.5 typical for soft materials.\cite{MichaelRubinstein_OU2003} We then computed the engineering stress from the diagonal elements of the measured stress tensor. As an example, for deformation in the $x$ direction, we extended the box as $\lambda=L_{x}/L_{x,0}=1+\dot\lambda t$, where $L_{x,0}$ and $L_x$ are the initial and extended box lengths in the $x$ direction, respectively. We calculated the engineering stress as $\sigma_{\mathrm{E}}=\left [ \sigma_{xx} - 0.5(\sigma_{yy}+\sigma_{zz}) \right ]/\lambda$, where $\sigma_{\alpha\alpha}$ is the diagonal element for direction $\alpha$ in the stress tensor. Typically, we simulated $\lambda$ in the range of 1.0 to 3.5 (time period $5\times 10^3\,\tau$). Since the extension rate can affect the tensile behavior of viscoelastic media, we repeated the uniaxial deformation simulations for the semiflexible system at half the extension rate (Fig.~S5). We found that a lower extension rate leads to a decrease in $\sigma_{\mathrm{E}}$, consistent with previous studies,\mbox{\cite{hossainMolecularDynamicsSimulations2010, lvMolecularDynamicsSimulation2021}} but does not significantly alter the overall trend of the stress curve or the corresponding mechanical properties. Moreover, in most cases, we found consistent initial behavior between independent simulations and extension directions (Fig.~S6), so we averaged them together to improve statistics.

\begin{figure}
\centering
\includegraphics{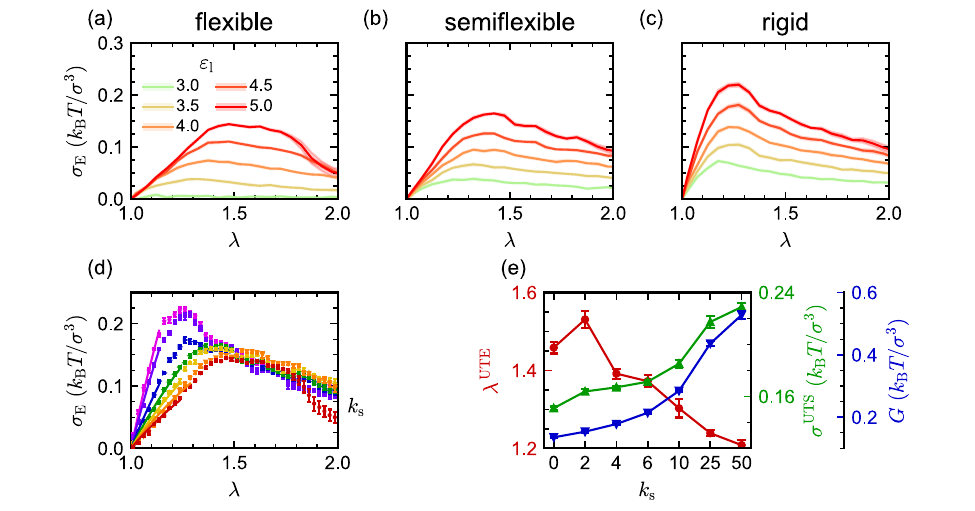}
\caption{Engineering stress $\sigma_{\mathrm{E}}$ for ligand-grafted nanoparticle assemblies under uniaxial tensile deformation (extension factor $\lambda$) for (a) flexible, (b) semiflexible, and (c) rigid ligands with varied attraction strength $\varepsilon_\mathrm{l}$. (d) The same as (a)--(c) for $\varepsilon_{\rm l} = 5$ and varied bending stiffness $k_{\rm s}$. The solid lines are the linear fits used to obtain the shear modulus $G$ for small deformations. (e) Shear modulus $G$, ultimate tensile strength $\sigma^{\mathrm{UTS}}$, and its corresponding ultimate tensile extension factor $\lambda^{\mathrm{UTE}}$ as functions of bending stiffness $k_{\mathrm{s}}$ for $\varepsilon_\mathrm{l}=5$, extracted from (a)--(d). Error bars represent the standard error of the mean across independent simulations and extension directions.}
\label{figure3}
\end{figure}

We found that $\sigma_{\mathrm{E}}$ gradually increased for small deformations for ligand stiffnesses $k_{\mathrm{s}}\lesssim 6$, corresponding to flexible and semiflexible ligands [Figs.~\ref{figure3}(a)--(b) and S7]. For rigid ligands, $\sigma_{\mathrm{E}}$ increased more rapidly with small changes in $\lambda$ for all $\varepsilon_\mathrm{l}$ values, and also exhibited a narrower peak at smaller $\lambda$ values compared to the flexible and semiflexible ligands [Figs.~\ref{figure3}(c) and S7]. This behavior is most apparent when $\sigma_{\rm E}$ is compared for different bending stiffnesses $k_{\rm s}$ at the strongest ligand--ligand attraction simulated [$\varepsilon_{\rm l} = 5$, Fig.~\ref{figure3}(d)]. To further compare mechanical properties as a function of ligand flexibility, we extracted the maximum value of $\sigma_{\mathrm{E}}$ as the ultimate tensile strength $\sigma^{\mathrm{UTS}}$, and its corresponding ultimate tensile extension factor $\lambda^{\mathrm{UTE}}$ as an indicator of network stretchability from Fig.~\ref{figure3}(d). Following a recent study,\cite{MurariSingh_JPCC2022} we also determined the shear modulus $G$, which quantifies the rigidity of the network, by fitting $\sigma_{\mathrm{E}}=G(\lambda-1/\lambda^{2})$ in the limit of small deformations [fitted regions shown in Fig.~\ref{figure3}(d)]. We found that $G$ and $\sigma^{\mathrm{UTS}}$ increased moderately with increasing $k_{\mathrm{s}}$ up to $k_{\mathrm{s}}=6$ (\textit{i.e.}, for flexible and semiflexible ligands), and then significantly increased for more rigid ligands [Fig.~\ref{figure3}(e)]. In contrast, $\lambda^{\mathrm{UTE}}$ showed a distinct transition: it remained around $\lambda^{\mathrm{UTE}} = 1.4$ for $k_{\mathrm{s}} \leq 6$, but sharply decreased to approximately $\lambda^{\mathrm{UTE}} = 1.2$ for more rigid ligands. We attribute this reduction to the more limited internal molecular degrees of freedom of rigid ligands [Fig.~\ref{figure3}(e)]. These observations highlight that flexible and semiflexible ligands produced soft, weak nanoparticle gels that were more stretchable than those produced using rigid ligands.
\subsection{Comparison to linker-mediated nanoparticle assemblies} \label{Sec3.2}
\begin{figure}
\centering
\includegraphics{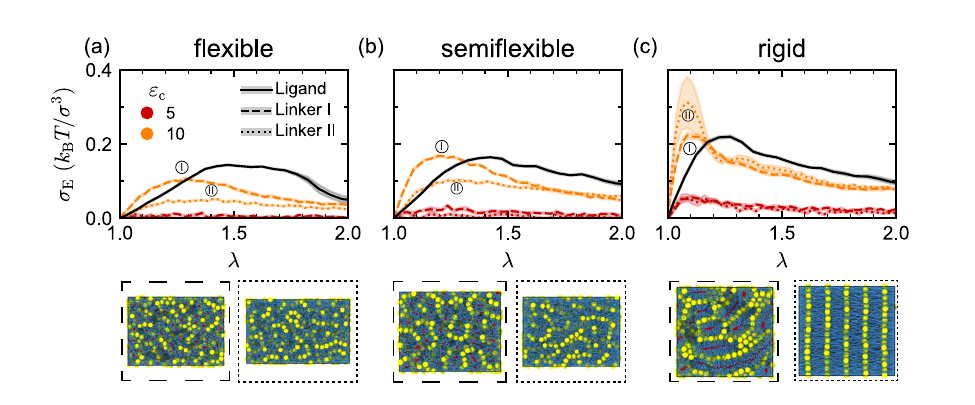}
\caption{Engineering stress $\sigma_{\mathrm{E}}$ response of Linker I-mediated (dashed lines) and Linker II-mediated (dotted lines) nanoparticle assemblies to uniaxial tensile deformation (extension factor $\lambda$) at a fixed linker--linker attraction strength $\varepsilon_\mathrm{l}=5$ and two different linker--nanoparticle attraction strengths $\varepsilon_\mathrm{c}$ for (a) flexible, (b) semiflexible, and (c) rigid chains. The solid lines in (a)--(c) show results for the ligand-grafted nanoparticle assemblies with ligand--ligand attraction strength $\varepsilon_\mathrm{l}=5$. Simulation snapshots corresponding to the maximum in $\sigma_{\mathrm{E}}$ for Linker I and Linker II are shown below each panel. Error bars represent the standard error of the mean across independent simulations and extension directions.}
\label{figure4}
\end{figure}
After characterizing how the mesostructure and mechanical properties of the ligand-grafted nanoparticle assemblies varied with ligand--ligand attraction and ligand bending stiffness, we proceeded to investigate how this behavior was similar or different from that in linker-mediated nanoparticle assemblies. As described in Section \ref{Sec2}, we studied two different types of linkers: Linker I, having one end that facilitates linker--linker attraction and one end that facilitates linker--nanoparticle attraction, and Linker II, which is twice as long as Linker I and whose ends facilitate only linker--nanoparticle attraction. For Linker I, we fixed the linker--linker attraction strength to $\varepsilon_\mathrm{l}=5$ to compare with the most attractive ligand-grafted nanoparticles. Linker II does not have this parameter because it is effectively two Linker I molecules that are permanently associated together. We used these linkers to create nanoparticle assemblies and compute their mechanical properties for different linker--nanoparticle attraction strengths $\varepsilon_\mathrm{c}$ at the same volume fraction ($\eta = 0.13$) that was used for the ligand-grafted nanoparticles.

We computed the average engineering stress $\sigma_{\mathrm{E}}$ by applying uniaxial tensile deformation on the assemblies formed with Linker I and Linker II for varying linker stiffness at two attraction strengths $\varepsilon_\mathrm{c} = 5\,\textrm{and}\,10$ [Figs.~\ref{figure4}(a)--(c) and S8]. We found that the rate of change in $\sigma_{\mathrm{E}}$ at small deformations increased monotonically with increasing linker stiffness for both Linker I and Linker II. In contrast, ligand-mediated nanoparticle assemblies only exhibited prominent differences in the stress response at small deformations when the ligands became rigid, as shown by comparing the stress response for the ligand- and linker-mediated assemblies at $\varepsilon_\mathrm{c}=10$ [Figs.~\ref{figure4}(a)--(c) and S8]. Further, whether a maximum in $\sigma_{\mathrm{E}}$ was attained using Linker I or Linker II depended on the chain stiffness.

One direct cause of these differences in mechanical properties could be variation in association resulting from attraction or bond formation between beads. To clarify the association of terminal beads,, we computed various local coordination metrics for the nanoparticles and polymers in the different models. Figure S9(a) shows that when $\varepsilon_\mathrm{c} = 5$, a significant number of linkers are not attached to the particle surface, but as $\varepsilon_\mathrm{c}$ increases to $10$, such free linkers are almost absent. This is also reflected in the weak $\sigma_{\mathrm{E}}$ observed for $\varepsilon_\mathrm{c} = 5$, whereas at $\varepsilon_\mathrm{c} = 10$, the tensile behavior becomes comparable to that of ligand-mediated nanoparticle assemblies [Figs.~4(a)--(c) and S8]. We further examined the specificity of the attraction between linker terminals and surface sites. Figure~S9(b) shows that the average number of linkers surrounding a nanoparticle exceeds 14, which is the number of ligands per nanoparticle, indicating that the distribution of linkers and ligands onto nanoparticles also differs.

We also visually noted that the linker-attaching terminal beads of Linker I tended to cluster (Fig.~4), so we quantified this aggregation and compared the binding of the terminal beads in the Linker I and ligand models (Fig.~S10). At $\varepsilon_\mathrm{l} = 5$, the linker-attaching terminal beads of both models were mostly bound to each other [Fig.~S10(a)], and their average coordination number was around 7 [Fig.~S10(b)]. Notably, in the semiflexible regime, the difference in terminal bead aggregation between the ligand and linkers [Fig.~S9(b)] or between the two types of linkers [Fig.~S10(b)] is minimal; however, their $\sigma_{\mathrm{E}}$ responses differ significantly [Fig.~4(b)], suggesting that the mechanical behavior of the gels is likely influenced by the overall network structure rather than just the aggregation of the terminal beads.

Meanwhile, we also noticed significant variability in the stress curves of rigid Linker II, which we attribute to the formation of lamellar-like structures [Fig.~4(c)]. Once the lamellae form, the mechanical properties become highly directional (Fig.~S11). In the direction perpendicular to the lamellae, the $\sigma^{\mathrm{UTS}}$ increases significantly. Because the orientation of the lamellae is random, isolating the effects of anisotropy in these structures is challenging. Therefore, in the subsequent analysis, we continue to average across all extension directions, which might be regarded as an orientational average.
\begin{figure}
\centering
\includegraphics{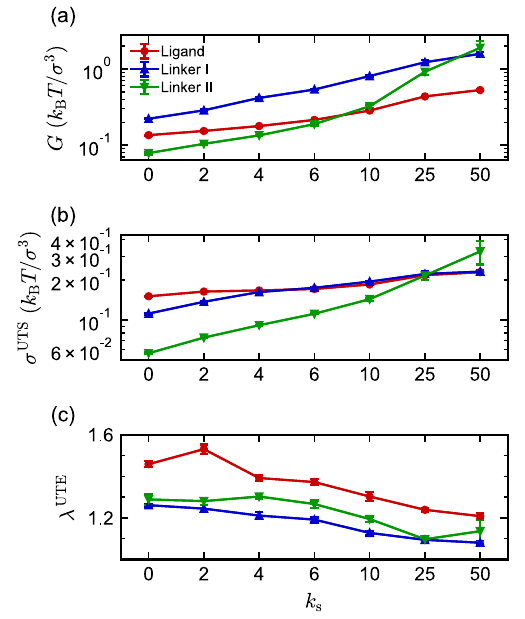}
\caption{(a) Shear modulus $G$, (b) ultimate tensile strength $\sigma^{\mathrm{UTS}}$, and (c) its corresponding strain $\lambda^{\mathrm{UTE}}$ of gels as functions of bending stiffness $k_{\mathrm{s}}$ for ligand-grafted (red), Linker I mediated (blue), and Linker II mediated (green) nanoparticle assemblies. These mechanical properties are extracted from the strain-stress curves shown in Figs.~\ref{figure3} and \ref{figure4}, respectively. For the linked nanoparticle assemblies, the properties are shown for linker-nanoparticle attraction strength of $\varepsilon_\mathrm{c}=10$. Error bars represent the standard error of the mean across independent simulations and extension directions.}
\label{figure5}
\end{figure}
To better quantify the impact of chain stiffness and association scheme on mechanical properties, we computed $\sigma^{\mathrm{UTS}}$, $\lambda^{\mathrm{UTE}}$, and $G$ for the linker-mediated nanoparticle assemblies. To compare between the different models, we focused on $\varepsilon_\mathrm{c}=10$, which yielded similar magnitudes of stress for various bending stiffnesses [Figs.~\ref{figure5}(a)--(c)]. We found that Linker I, despite having the same length and self-interactions as the ligands, produced nanoparticle assemblies with distinct mechanical properties: the assemblies were more rigid (\textit{i.e.}, higher $G$) and less stretchable (\textit{i.e.}, lower $\lambda^{\mathrm{UTE}}$) compared to the ligand-mediated nanoparticle assemblies, while exhibiting similar ultimate tensile strength $\sigma^{\mathrm{UTS}}$ at all $k_{\mathrm{s}}$ values. As we have shown above, the aggregation behavior of their linker-attaching beads is similar [Fig.~S9(b)]. Therefore, we attribute the differences in $G$ and $\lambda^{\mathrm{UTE}}$ to the tendency of the nanoparticles to form clusters in the linker-mediated nanoparticle assemblies\cite{MurariSingh_JPCC2022} (Fig.~S12 and Movies S1--S6), as well as the disruption of attractions between surface sites and nanoparticle-attaching beads (Fig.~S13). When flexible and semiflexible Linker II (\textit{i.e.}, $k_{\mathrm{s}} \lesssim 6$) were used, the resulting assemblies were softer (lower $G$) and weaker (lower $\sigma^{\mathrm{UTS}}$) compared to both the Linker I and ligand-mediated assemblies, but the extent to which they could be stretched, as indicated by $\lambda^{\mathrm{UTE}}$, fell in between that of Linker I and the ligands, most likely because of the longer chain length of Linker II. This variation in extension capability may also be related to the dispersion of the nanoparticles. Visualization showed that the nanoparticles with linkers tended to locally aggregate (Movies S1--S9), reducing the uniformity of the percolated network, which we hypothesize may make it more susceptible to deformation than ligand-mediated networks. For $k_{\mathrm{s}} > 10$, both $\sigma^{\mathrm{UTS}}$ and $G$ of the Linker II systems increased rapidly due to the emergence of lamellar-like structural ordering, which was also highlighted by the presence of sharp peaks in $S(q)$ (Figs.~S14 and S15). This lamellar structure formed when there was sufficient attraction ($\varepsilon_\mathrm{c} \geq 4$) between the rigid Linker II chains and the nanoparticles (Fig.~S16). Such ordered structures can exhibit high resistance to tensile deformation before rupturing,\cite{JihoonChoi_JACS2010} which is consistent with the high values of $\sigma^{\mathrm{UTS}}$ and $G$ for rigid Linker II (also see Movies S1--S9).

\section{Conclusions}
In this computational study, we investigated the phase behavior and mechanical properties of gel-like nanoparticle assemblies produced using associative polymers as either grafted ligands or free linkers. In the ligand-mediated nanoparticle assemblies, we found that sufficiently strong ligand--ligand attractions produced percolated networks over a broad range of nanoparticle volume fractions that gradually narrowed with increasing ligand stiffness. Under uniaxial tensile deformation, the strength and rigidity of these networks increased with increasing ligand stiffness, while the network stretchability decreased. When we replaced the ligands with comparable free linkers, the resulting nanoparticle assemblies displayed distinctly different mechanical properties. The linker-mediated nanoparticle assemblies were more rigid and less stretchable, irrespective of the bending stiffness, owing to the closely packed arrangement of nanoparticles. Nevertheless, both linker- and ligand-mediated nanoparticle assemblies exhibited comparable tensile strength. When the linker length was doubled and association occurred only through linker--nanoparticle attractions, the assemblies became considerably softer, weaker, and less stretchable compared to the ligand-mediated assemblies. Our findings provide molecular-level insights into how different polymer association schemes regulate, and hence may be used to tune, the thermodynamics and mechanical response of nanoparticle gels.

\section*{Data Availability}
The data that supports the findings of this study are available from the authors upon reasonable request.

\section*{Conflicts of Interest}
There are no conflicts to declare.

\section*{Acknowledgments}
Research was sponsored by the Army Research Office and was accomplished under Grant Number W911NF-24-1-0245 (to JM). The conclusions contained in this document are those of the authors and should not be interpreted as representing the official policies, either expressed or implied, of the Army Research Office or the U.S. Government. The U.S. Government is authorized to reproduce and distribute reprints for Government purposes notwithstanding any copyright notation herein. JM also acknowledges the support provided by the Welch Foundation under the grant A-2113-20220331. MPH acknowledges support from the National Science Foundation under Award No.~2223084 and the International Fine Particle Research Institute. AN acknowledges funding by the Deutsche Forschungsgemeinschaft (DFG, German Research Foundation) through projects 451785257 and 470113688.  We are grateful to the Texas A{\upshape\&}M High Performance Research Computing (HPRC) for providing the computational resources required to complete this work.

\bibliography{ref_Abb}
\bibliographystyle{vancouver}
\end{document}


\title
{Supplementary Information for
``Effects of Ligand vs.~Linker on Phase Behavior and Mechanical Properties of Nanoparticle Gels''}

\author{Qizan Chen}
\affiliation{Artie McFerrin Department of Chemical Engineering, Texas A{\upshape\&}M University, College Station, TX 77843, United States}

\author{Dinesh Sundaravadivelu Devarajan}
\affiliation{Artie McFerrin Department of Chemical Engineering, Texas A{\upshape\&}M University, College Station, TX 77843, United States}

\author{Arash Nikoubashman}
\affiliation{Leibniz-Institut f{\"u}r Polymerforschung Dresden e.V., Hohe Stra{\ss}e 6, 01069 Dresden, Germany}
\affiliation{Institut f{\"u}r Theoretische Physik, Technische Universit{\"a}t Dresden, 01069 Dresden, Germany}

\author{Michael P. Howard}
\email{mphoward@auburn.edu}
\affiliation{Department of Chemical Engineering, Auburn University, Auburn, Alabama 36849, United States}

\author{Jeetain Mittal}
\email{jeetain@tamu.edu}
\affiliation{Artie McFerrin Department of Chemical Engineering, Texas A{\upshape\&}M University, College Station, TX 77843, United States}
\affiliation{Department of Chemistry, Texas A{\upshape\&}M University, College Station, TX 77843, United States}
\affiliation{Interdisciplinary Graduate Program in Genetics and Genomics, Texas A{\upshape\&}M University, College Station, TX 77843, United States}

\maketitle

\FloatBarrier
\section{Supplementary Figures} \label{Sec1}
\beginsupplement

\begin{figure}[H]
\centering
\includegraphics{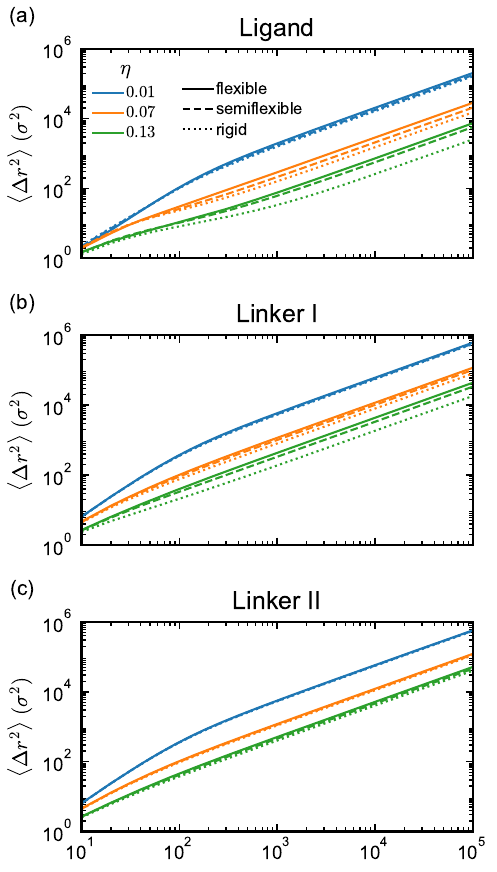}
\caption{Mean squared displacement of (a) ligand-, (b) Linker I-, and (c) Linker II-mediated nanoparticles with purely repulsive interactions. Colors indicate different volume fractions $\eta$, and line styles represent polymer bending stiffness: solid (flexible), dashed (semiflexible), and dotted (rigid).Simulations were run for $5\times{10}^5\tau$ , starting from the initial dispersed state prior to annealing, with nanoparticle centers saved every $10\tau$}
\label{figureS1}
\end{figure}

\begin{figure}
\centering
\includegraphics{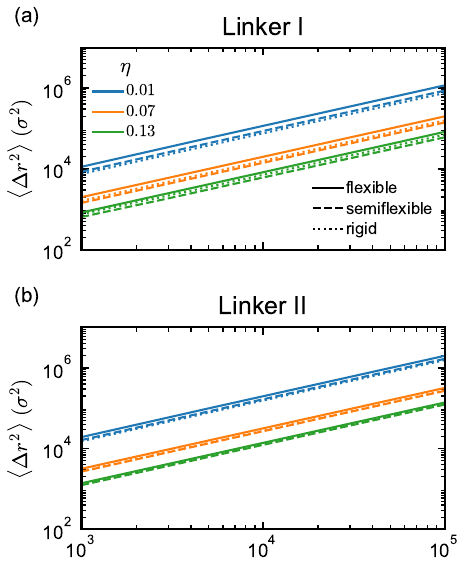}
\caption{Mean squared displacement of polymer monomers in (a) Linker I- and (b) Linker II-mediated nanoparticles with purely repulsive interactions. Colors indicate different volume fractions $\eta$, and line styles represent polymer bending stiffness: solid (flexible), dashed (semiflexible), and dotted (rigid).The simulations and analysis protocols are identical to those described for Fig.~S1; except that polymer monomer positions were saved every $10^{3}\tau$ to reduce storage demands.}
\label{figureS2}
\end{figure}

\begin{figure}
\centering
\includegraphics{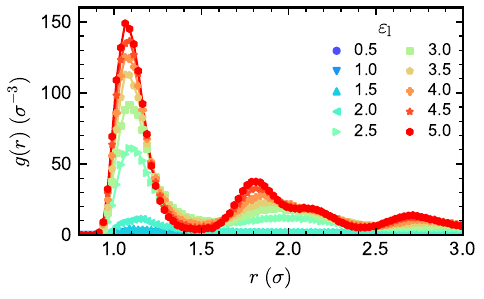}
\caption{Radial distribution function between attractive beads of flexible ligands on different nanoparticles as a function of ligand--ligand attraction strength $\varepsilon_\mathrm{l}$.}
\label{figureS3}
\end{figure}

\begin{figure}
\centering
\includegraphics{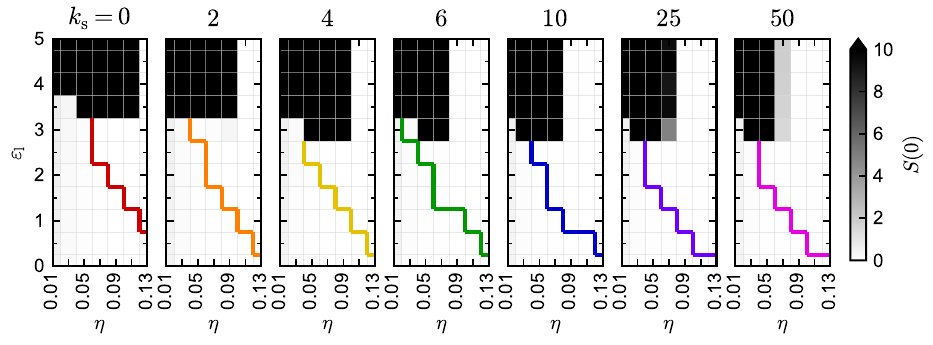}
\caption{Nanoparticle partial static structure factor extrapolated to zero wavevector, $S(0)$, and percolation line as a function of volume fraction $\eta$ and ligand--ligand attraction strength $\varepsilon_\mathrm{l}$ for different ligand bending stiffnesses $k_{\rm s}$.}
\label{figureS4}
\end{figure}

\begin{figure}
\centering
\includegraphics{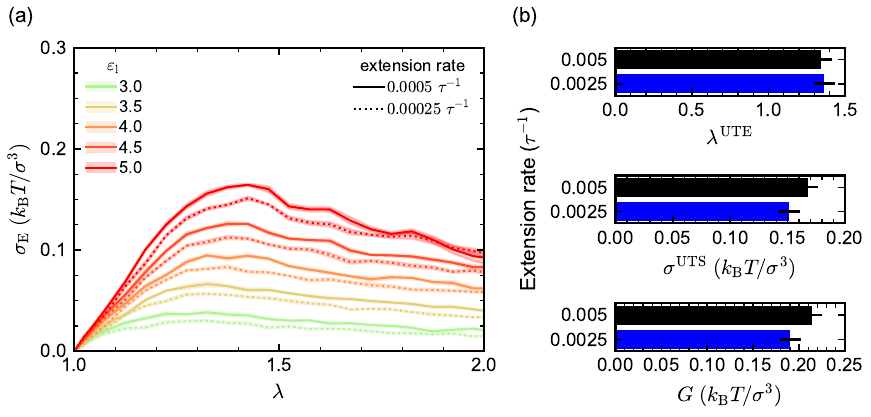}
\caption{(a) Engineering stress $\sigma_{\rm E}$ as a function of strain \(\lambda\) for semiflexible ligands under two extension rates \(0.00025 \tau^{-1}\) (dashed lines) and \(0.0005 \tau^{-1}\) (solid lines). (b) Comparison of mechanical properties \(\lambda^{\mathrm{UTE}}\), \(\sigma^\mathrm{UTS}\), and \(G\) for \(\varepsilon_\mathrm{l} = 5\) at the two extension rates. Error bars represent the standard error of the mean across independent simulations and extension directions.}
\label{figureS5}
\end{figure}

\begin{figure}
\centering
\includegraphics{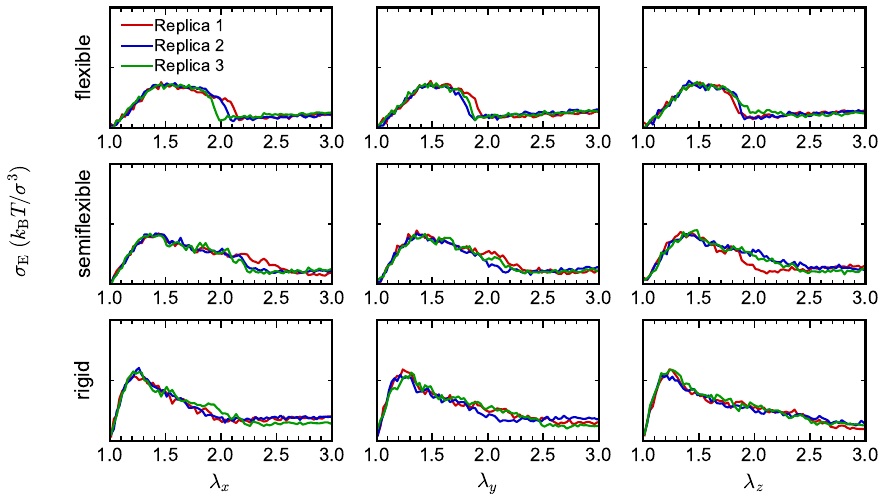}
\caption{Engineering stress $\sigma_{\mathrm{E}}$ for ligand-mediated nanoparticle assemblies under uniaxial tensile deformation (extension factor $\lambda$) in three different stretching directions and three independent simulations for flexible, semiflexible, and rigid ligands.}
\label{figureS6}
\end{figure}

\begin{figure}
\centering
\includegraphics{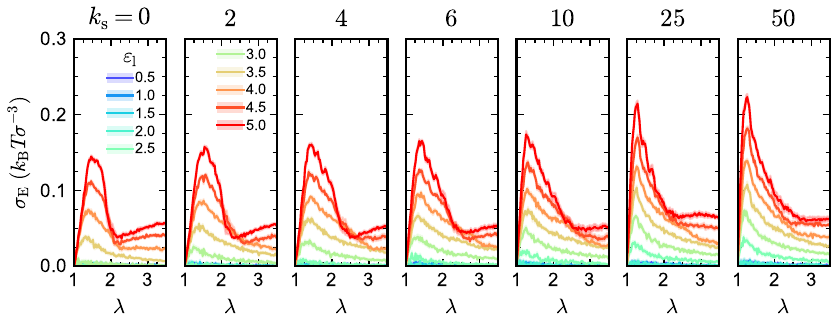}
\caption{Engineering stress $\sigma_{\mathrm{E}}$ for ligand-mediated nanoparticle assemblies under uniaxial tensile deformation (extension factor $\lambda$) for different ligand bending stiffnesses $k_{\rm s}$. Error bars represent the standard error of the mean across independent simulations and extension directions.}
\label{figureS7}
\end{figure}

\begin{figure}
\centering
\includegraphics{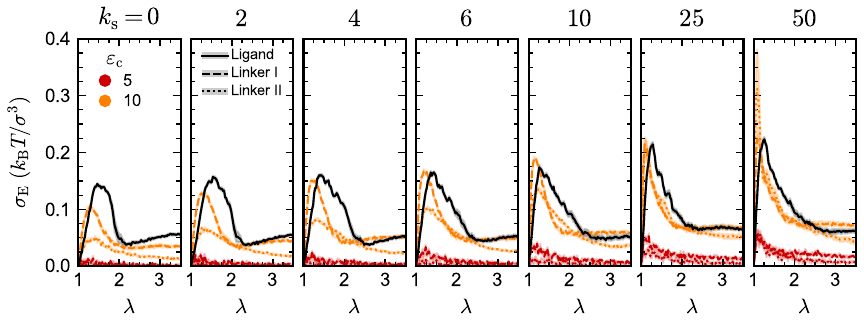}
\caption{Engineering stress $\sigma_{\mathrm{E}}$ of Linker I- (dashed lines) and Linker II- (dotted lines) mediated nanoparticle assemblies to uniaxial tensile deformation (extension factor $\lambda$) at a fixed linker--linker attraction strength $\varepsilon_\mathrm{l}=5$ and two different linker--nanoparticle attraction strengths $\varepsilon_\mathrm{c}$ for different linker bending stiffnesses $k_{\rm s}$. The solid lines show the same measurements for the ligand-mediated nanoparticle assemblies with ligand--ligand attraction strength $\varepsilon_\mathrm{l}=5$. Error bars represent the standard error of the mean across independent simulations and extension directions.}
\label{figureS8}
\end{figure}

\begin{figure}
\centering
\includegraphics{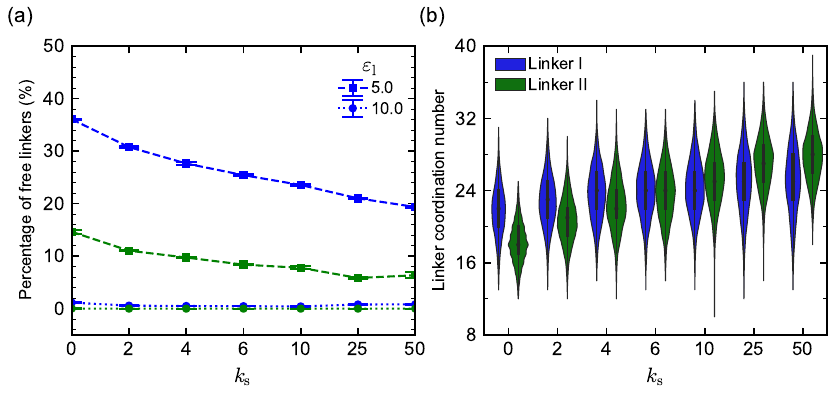}
\caption{(a) Percentage of free linkers as a function of polymer bending stiffness $k_{\mathrm{s}}$ for Linker I- (blue) and Linker II- (green) mediated nanoparticle assemblies with $\eta = 0.13$, $\varepsilon_{\mathrm{l}} = 5$, and varied $\varepsilon_{\mathrm{c}} = 5$ or $10$. Error bars represent the standard error of the mean across the final assemblies from annealing. (b) Corresponding violin plot of linker coordination number per nanoparticle at $\varepsilon_{\mathrm{c}} = 10$.}
\label{figureS9}
\end{figure}

\begin{figure}
\centering
\includegraphics{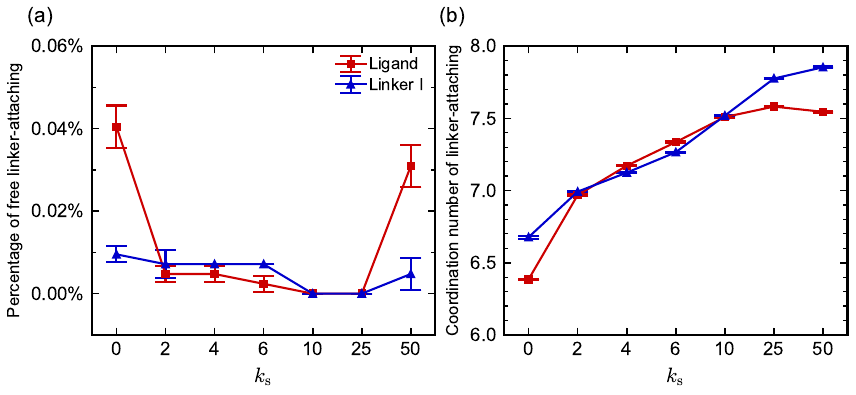}
\caption{(a) Percentage of free linker-attaching beads and (b) averaged linker-attaching coordination number as a function of polymer bending stiffness $k_{\mathrm{s}}$ for ligand- (red) and Linker I- (blue) mediated nanoparticle assemblies with $\eta = 0.13$, $\varepsilon_{\mathrm{l}} = 5$ and $\varepsilon_{\mathrm{c}} = 10$. Error bars represent the standard error of the mean across the final assemblies from annealing.}
\label{figureS10}
\end{figure}

\begin{figure}
\centering
\includegraphics{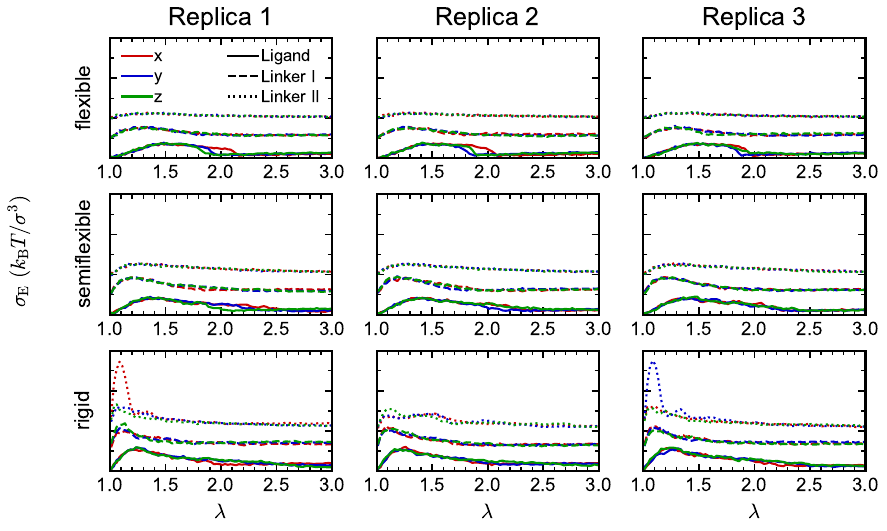}
\caption{Engineering stress $\sigma_{\rm E}$ as a function of the extension factor $\lambda$ in the $x$, $y$, and $z$ directions for ligand- (solid Line), Linker I- (dashed Line), and Linker II- (dotted Line) mediated nanoparticle assemblies using flexible, semiflexible, and rigid polymers. Results are shown for three independent simulations.}
\label{figureS11}
\end{figure}

\begin{figure}
\centering
\includegraphics{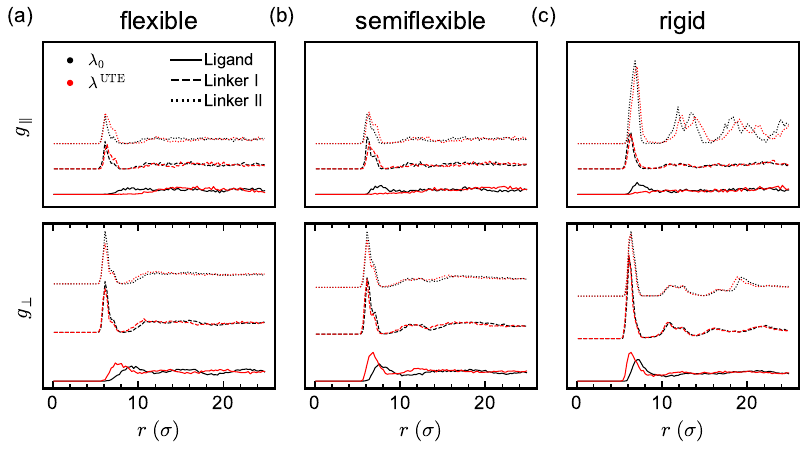}
\caption{Nanoparticle--nanoparticle radial distribution function parallel $g_{||}$ and perpendicular $g_\bot$ to the direction of deformation for (a) flexible, (b) semiflexible, and (c) rigid ligand, Linker I and Linker II polymers. The black and red lines correspond to systems before ($\lambda=0$) and after deforming to $\lambda=\lambda^{\mathrm{UTE}}$, respectively. The calculation was done following the methodology described in Ref.~7.}
\label{figureS12}
\end{figure}

\begin{figure}
\centering
\includegraphics{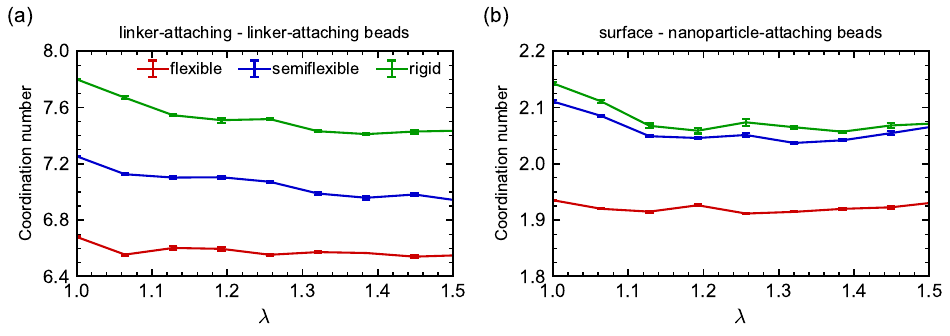}
\caption{Coordination number of beads as a function of the extension factor $\lambda$ for flexible, semiflexible and rigid Linker I-mediated nanoparticle assemblies with $\varepsilon_\mathrm{l}=5$ and $\varepsilon_\mathrm{c}=10$. (a) Coordination number between linker-attaching beads. (b) Coordination number between surface and nanoparticle-attaching beads. Error bars represent the standard error of the mean across independent simulations and extension directions.}
\label{figureS13}
\end{figure}

\begin{figure}
\centering
\includegraphics{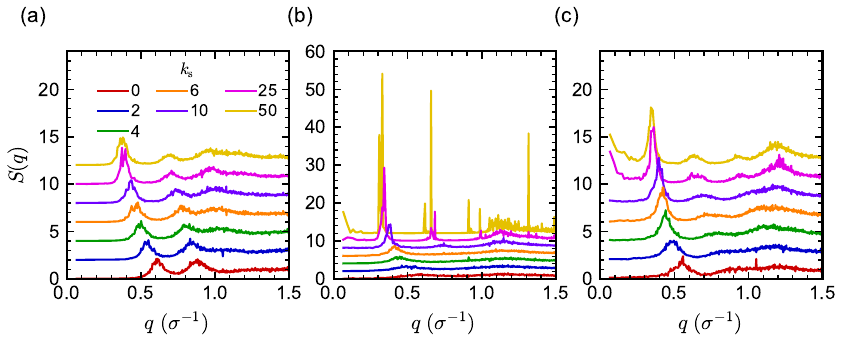}
\caption{Nanoparticle partial static structure factor $S(q)$ for different bending stiffnesses $k_{\rm s}$: (a) ligand-mediated nanoparticle assemblies with ligand--ligand attraction strength $\varepsilon_\mathrm{l}=5$, (b) Linker I-mediated nanoparticle assemblies with linker--linker attraction strength $\varepsilon_\mathrm{l}=5$ and linker--nanoparticle attraction strength  $\varepsilon_\mathrm{c}=10$, and (c) Linker II-mediated nanoparticle assemblies with linker--nanoparticle attraction strength  $\varepsilon_\mathrm{c}=10$. The data lines are additionally offset by 2 in order from smallest to largest $k_{\rm s}$ to better compare the differences.}
\label{figureS14}
\end{figure}

\begin{figure}
\centering
\includegraphics{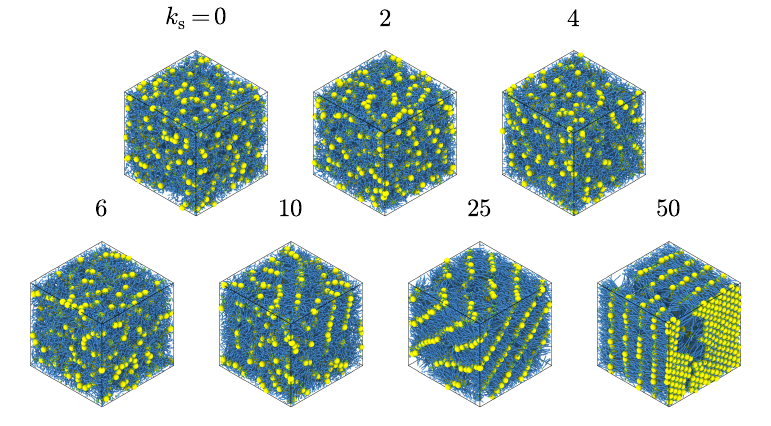}
\caption{\justifying  Simulation snapshots of Linker II-mediated nanoparticle assemblies with a fixed linker--nanoparticle attraction strength $\varepsilon_\mathrm{c}=10$ for all bending stiffnesses $k_{\rm s}$ investigated.}
\label{figureS15}
\end{figure}

\begin{figure}
\centering
\includegraphics{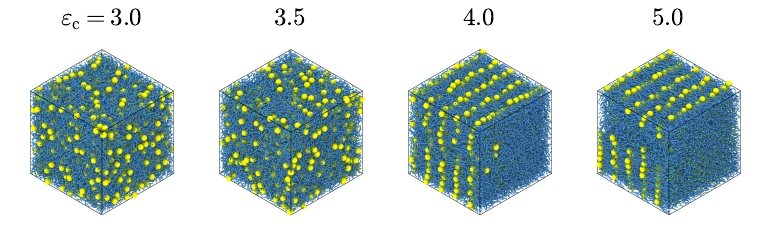}
\caption{\justifying Simulation snapshots of Linker II-mediated nanoparticle assemblies with fixed bending stiffness $k_{\rm s}=50$ as a function of linker--nanoparticle attraction strength $\varepsilon_\mathrm{c}$.}
\label{figureS16}
\end{figure}

\FloatBarrier
\section{Supplementary Movies} \label{Sec2}

\begin{itemize}
\item Movie S1. Uniaxial tensile deformation of flexible ligand-mediated nanoparticle assembly for ligand--ligand attraction strength $\varepsilon_\mathrm{l}=5$.

\item Movie S2. Uniaxial tensile deformation of semiflexible ligand-mediated nanoparticle assembly for ligand--ligand attraction strength $\varepsilon_\mathrm{l}=5$.

\item Movie S3. Uniaxial tensile deformation of rigid ligand-mediated nanoparticle assembly for ligand--ligand attraction strength $\varepsilon_\mathrm{l}=5$.

\item Movie S4. Uniaxial tensile deformation of flexible Linker I-mediated nanoparticle assembly for linker--linker attraction strength $\varepsilon_\mathrm{l}=5$ and linker--nanoparticle attraction strength $\varepsilon_\mathrm{c}=10$.

\item Movie S5. Uniaxial tensile deformation of semiflexible Linker I-mediated nanoparticle assembly for linker--linker attraction strength $\varepsilon_\mathrm{l}=5$ and linker--nanoparticle attraction strength $\varepsilon_\mathrm{c}=10$.

\item Movie S6. Uniaxial tensile deformation of rigid Linker I-mediated nanoparticle assembly for linker--linker attraction strength $\varepsilon_\mathrm{l}=5$ and linker--nanoparticle attraction strength $\varepsilon_\mathrm{c}=10$.

\item Movie S7. Uniaxial tensile deformation of flexible Linker II-mediated nanoparticle assembly for linker--nanoparticle attraction strength $\varepsilon_\mathrm{c}=10$.

\item Movie S8. Uniaxial tensile simulation of semiflexible Linker II-mediated nanoparticle assembly for linker--nanoparticle attraction strength $\varepsilon_\mathrm{c}=10$.

\item Movie S9. Uniaxial tensile deformation of rigid Linker II-mediated nanoparticle assembly for linker--nanoparticle attraction strength $\varepsilon_\mathrm{c}=10$.
\end{itemize}
